\documentclass{epl}

\usepackage{graphicx,epsf,epsfig}
\usepackage[T1]{fontenc}

\title{Scaling approach to the phase diagram of quantum Hall systems}
\shorttitle{Scaling approach to the phase diagram of QH systems}
\author{M.\ O.\ Goerbig\inst{1,2} and C.\ Morais\ Smith\inst{1}}
\shortauthor{M. O. Goerbig \etal}

\institute{
\inst{1} D\'epartement de Physique, Universit\'e de Fribourg, P\'erolles,  CH-1700 Fribourg, Switzerland.\\
\inst{2} Laboratoire de Physique des Solides, Universit\'e Paris-Sud, F-91405 Orsay, France.}


\begin{document}

\maketitle

\begin{abstract}
We present a simple classification of the different liquid and solid
phases of quantum Hall systems in the regime where the Coulomb
interaction between electrons is significant, {\sl i.e.} away from integral 
filling factors. This classification,
and a criterion for the validity of the mean-field approximation in
the charge-density-wave phase, is based on scaling arguments concerning
the effective interaction potential of electrons restricted to an
arbitrary Landau level. Finite-temperature effects are investigated
within the same formalism, and 
a good agreement with recent experiments is obtained.
\end{abstract}

Two-dimensional electron systems (2DES) in a perpendicular magnetic
field exhibit a rich variety of phases, ranging from incompressible
quantum liquids, which are responsible for the integral and fractional
quantum Hall effects (IQHE and FQHE), to electron-solid phases such
as charge density waves (CDWs) and the Wigner
crystal (WC). The IQHE is found when the electron density $n_{el}$ is
an integral multiple of the density of states per Landau level (LL)
$n_B=1/2\pi l_B^2$, where $l_B=\sqrt{\hbar/eB}$ is the magnetic length
\cite{KvK}, and can be described within a single-particle picture. On
the other hand, when the filling factor $\nu=n_{el}/n_B$ is
non-integral, the highest filled LL has only a partial filling
$\bar{\nu}=\nu-n$, and it becomes essential to include the Coulomb interaction \cite{note1}.
It lifts the LL degeneracy and leads to a rich phase diagram. At
extremely low electron densities, an insulating phase has been 
observed whose properties are attributed to WC formation \cite{jiang}. At 
$\nu=p/q$, with $p,q$ integral and $q$ odd, the FQHE is observed in the two 
lowest LLs \cite{TSG}. The corresponding ground state is an incompressible liquid, 
analogous to the IQHE if described in terms of composite fermions (CFs)
\cite{jain}. At $\bar{\nu}= 1/2$ in higher LLs, huge anisotropies
have been detected in the longitudinal magnetoresistance, indicating the formation
of a unidirectional CDW \cite{exp}. From the theoretical point of
view, Hartree-Fock calculations predict correctly a CDW
formation around $\bar{\nu}=1/2$ in higher LLs \cite{FKS,MC}, but have
failed to describe the FQHE regime. In this letter we propose
a new scaling approach, which classifies the liquid and solid phases observed 
in quantum Hall systems in terms of length scales, and remains valid in all 
LLs. 
In addition, we clarify the reasons for breakdown of the Hartree-Fock approximation in the lowest LLs and establish a criterion for the appearance of recently observed FQHE features in the WC regime at finite temperatures \cite{pan}.

Because in the high-magnetic-field limit the Coulomb interaction between the electrons constitutes a smaller energy scale than the gap between LLs, inter-LL excitations are high-energy degrees of freedom. At integral filling, they are the only excitations, but may be neglected when considering a system at $\nu\neq n$ because in this case low-energy excitations become possible within the same LL \cite{AG}. The kinetic energy of the electrons may therefore be set to zero, and the Coulomb interaction remains as the only energy scale in the problem if the electrostatic potential due to underlying impurities is small. One obtains a system of strongly correlated electrons described by the Hamiltonian
\begin{equation}
\label{equ001}
\hat{H}_n=\frac{1}{2}\sum_{\bf q}v(q)\left[F_n(q)\right]^2\bar{\rho}(-{\bf q})\bar{\rho}({\bf q}),
\end{equation}
where $v(q)=2\pi e^2/\epsilon q$ is the two-dimensional Fourier transform of the Coulomb interaction. In this model, one considers only interactions between spinless electrons within the $n$th LL described by the density operators
$\langle \rho({\bf q})\rangle_n=F_n(q) \bar{\rho}({\bf q})$,
where $\rho({\bf q})$ is the usual electron density in reciprocal
space. The factors $F_n(q)=L_n(q^2/2)\exp(-q^2/4)$, with the Laguerre
polynomials $L_n(x)$, arise from the wave functions of electrons in the $n$th LL and may be absorbed into an effective interaction potential $v_n(q)=v(q)[F_n(q)]^2$. The quantum nature of the problem is contained in unusual commutation relations for the electron density operators \cite{GMP},
\begin{equation}
\label{equ003}
[\bar{\rho}({\bf q}),\bar{\rho}({\bf k})]=2i\sin\left(\frac{({\bf q}\times{\bf k})_zl_B^2}{2}\right)\bar{\rho}({\bf q}+{\bf k}).
\end{equation}
The Hamiltonian (\ref{equ001}) together with the commutation relations (\ref{equ003}) defines the full model, which was used recently as a starting point for the description of the FQHE in the lowest LL \cite{MS} and for the formation of CDWs in higher LLs \cite{FKS,MC}. Note the formal equivalence between electrons in the lowest LL at $\nu=\nu_0$ and electrons in a higher LL at $\bar{\nu}=\nu_0$ within this model.

\begin{figure}
\epsfysize+5cm
\centerline{\epsffile{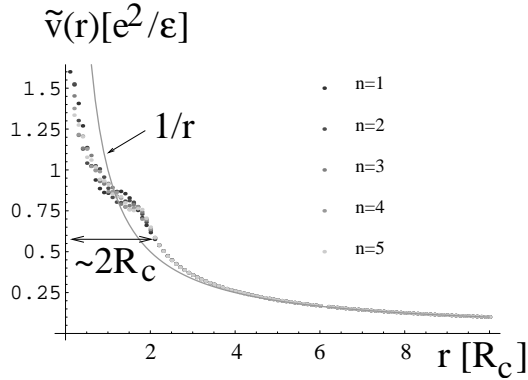}}
\caption{Effective interaction potential in real space. Points 
  correspond to $n=1,2,...,5$. 
  Gray line: pure Coulomb potential.}
\label{fig02}
\end{figure}

Deeper insight into the stucture of the model is obtained by 
transforming the effective interaction potential back to real space.
In appropriate units, one may derive a universal scaling function $\tilde{v}(r)$, 
\begin{equation}
\label{equ004}
v_n(r)=\sum_{\bf q}e^{-i{\bf q}\cdot{\bf r}}v_n(q)\approx\frac{\tilde{v}\left(r/R_C\right)}{\sqrt{2n+1}}\,,
\end{equation}
where $R_C=l_B\sqrt{2n+1}$ is the cyclotron radius. This scaling form becomes exact in the limit $n\rightarrow\infty$ because then $F_n(q)\rightarrow J_0(q\sqrt{2n+1})$. However, it is valid also at low values of $n$ as can be seen from fig.\,\ref{fig02}, where the rescaled results of $v_n(r)$ are shown for the LLs $n=1,...,5$.
The universal function exhibits a plateau of width
 $2R_C$ superimposed on the bare $1/r$ Coulomb potential, which 
is retrieved at large distances.
Although the bare Coulomb potential possesses no characteristic length scale, a component whose range is characterised by $R_C$ is introduced in the effective interaction potential. This permits a classification of the different phases according to the ratio of the average electronic separation $d\sim l_B/\sqrt{\bar{\nu}}$ in the $n$th LL and the range $2 R_C$ of the effective interaction.

\begin{figure}
\epsfysize+6.0cm
\centerline{\epsffile{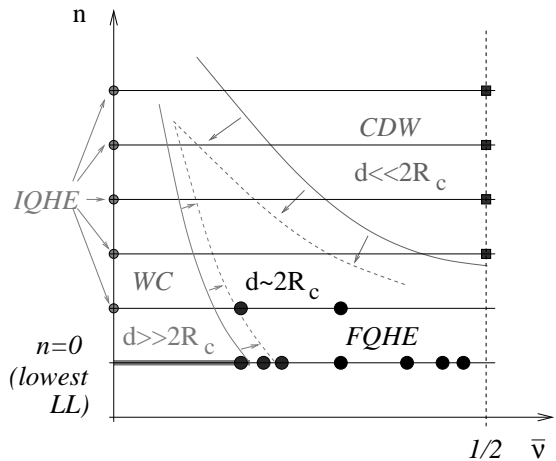}}
\caption{Phase diagram at $T=0$. Vertical and horizontal axes indicate, 
  respectively, the LL index $n$ and the partial filling $\bar{\nu}$. For 
  completely filled levels $\bar{\nu}=1$ one observes the IQHE (gray circles).
  The phase diagram is only shown up to half-filling $\bar{\nu}=1/2$ 
  because of particle-hole symmetry. 
  Black circles denote points at which the FQHE is found in experiments, the
  thick gray line shows the region of an experimentally observed insulating 
  phase attributed to WC formation \cite{jiang}, and gray squares indicate 
  points where the stripe phase is observed \cite{exp}. The full 
  lines represent the transition between different phases and move 
  towards the FQHE liquid phase in the presence of impurities (see 
  arrows and broken lines).}
\label{fig01}
\end{figure}

{\sl Quasi-classical limit (WC): $d\gg 2R_C$.} 
In the limit of low density, the
average distance between electrons interacting via the $1/r$ Coulomb potential is much larger than the spatial extent $R_C$ of their wave functions. Quantum corrections to the classical result are of order $\mathcal{O}(R_C/d)$ and may thus be neglected. Classically, the electrons are arranged in a triangular WC in order to minimise the repulsive Coulomb interaction. The transition line which separates the WC phase from other phases is obtained by comparing $d$ and $R_C$ as functions of $\bar{\nu}$ and $n$ (gray line in fig.\,\ref{fig01}). It is given by $\bar{\nu}_n^{WC}=\nu_0^{WC}/(2n+1)$, where $\nu_0^{WC}$ is the critical filling factor below which the WC is found in the lowest LL. Theoretical calculations predict $\nu_0^{WC}\approx 1/6.5$ for clean samples \cite{lam}, while experimentally the onset of WC behaviour is observed around $\nu=1/5$ \cite{jiang}.

Electrical transport in the WC phase arises either by a collective sliding 
mode or by the propagation of crystal dislocations. Sliding is suppressed 
by pinning of the WC due to residual impurities in the sample, and the 
number of dislocations is reduced by lowering the temperature. The 
experimental evidence for a WC phase arises from transport measurements 
\cite{jiang}, which indicate an insulating phase. In principle, this 
insulating behaviour could be attributed also to the 
localization of electrons by impurities \cite{AndersonLoc}. However, 
because the samples used for the measurements 
are extremely clean, this is unlikely to be the case.

{\sl Mean-field limit (CDWs): $d\ll 2R_C$.} 
Within the effective-potential framework, an arbitrarily chosen
particle interacts strongly with a number of neighbours
which can be estimated as $N_{n.n.}\approx \pi(2R_C)^2\bar{n}_{el}=
2(2n+1)\bar{\nu}$, where $\bar{n}_{el}$ is the density of electrons in the 
$n$th LL. A mean-field approximation, such as Hartree-Fock, 
is valid for large $N_{n.n.}$, where each particle may be considered
to interact only with an averaged background, without being influenced by the
individual motions of its neighbours. This limit cannot be obtained in 
the lowest LL, where the cyclotron radius coincides with the magnetic length
$l_B$. The average electronic separation would have to be shorter
than this length, which constitutes the smallest possible spacing because each
electronic state occupies a minimal surface $\sigma=2\pi l_B^2$. 
We note in addition that in the lowest LL the effective interaction 
potential does not exhibit a plateau as for $n\geq 1$.
The dark gray line in fig.\,\ref{fig01} shows the relation
$\bar{\nu}_n^{CDW}=N_c/2(2n+1)$. Comparison with experiment yields $N_{n.n.}=N_c\sim 5$ for the limiting value above which the mean-field approximation is justified \cite{exp}. 

The validity of the mean-field approximation in the regime $d\ll 2R_C$ also becomes apparent in Fourier space, where the wave vectors are renormalised in the same manner as the distances in eq.\,(\ref{equ004}). At large $n$, only the small-wave-vector limit remains important, and the commutation relation (\ref{equ003}) for the density operators becomes
\begin{eqnarray}
\nonumber
\left[\bar{\rho}\left(\frac{\bf q}{\sqrt{2n+1}}\right)\right.&,&\left.\bar{\rho}\left(\frac{\bf k}{\sqrt{2n+1}}\right)\right]\\
\nonumber
\approx &i& \frac{({\bf q}\times{\bf k})_zl_B^2}{2n+1}\ \bar{\rho}\left(\frac{{\bf q}+{\bf k}}{\sqrt{2n+1}}\right)=\mathcal{O}\left(\frac{1}{n}\right),
\end{eqnarray}
 after rescaling and  expansion of the sine function. The complicated algebraic structure of the density operators, and thus the quantum mechanical nature of the problem, become less important in higher LLs (larger $n$). This quasi-classical limit is different from the WC regime, in which the overlap of different electronic wave functions may be neglected. Here their overlap is sufficiently strong that the exchange interaction, which is included at the mean-field level, is essential.

The mean-field solution of the Hamiltonian (\ref{equ001}) reveals that the ground state in this limit is a CDW with a characteristic period on the order of the cyclotron radius $R_C$ \cite{FKS,MC}. This clustering of electrons, in spite of their repulsive interaction, may be understood qualitatively from the form of the effective interaction potential in fig.\,\ref{fig02}: if two electrons approach more than their average separation $d$, only a small additional energy cost is incurred because of the plateau in the region $r<2R_C$. However, a large energy on the order of the height of the plateau may be gained if the clustered electrons thus reduce the number of other electrons with which they interact strongly. The boundary above which the mean-field approximation becomes valid need not necessarily coincide with the CDW-FQHE phase transition. A detailed calculation of the ground-state energy would be needed to determine the exact transition line \cite{FKS,MC,fogler}, but the present scaling investigations serve as an upper limit for this boundary.

In the region $\bar{\nu}< 1/2$, charged clusters (or ``bubbles'') of several electrons form a super-WC to minimise their Coulomb repulsion (triangular CDW). Exactly at $\bar{\nu}=1/2$, however, these bubbles percolate to form lines and thus give rise to a ``stripe'' pattern (unidirectional CDW). This breaking of rotational symmetry arises because of a competing symmetry: the particle-hole symmetry becomes exact at half filling. A phase of bubbles would violate this symmetry, and a very small residual anisotropy in the underlying crystal suffices to fix the direction of the stripes \cite{scheidel}. However, there are other inhomogeneous charge configurations, which would also account for these symmetries, and there is theoretical evidence that a stripe pattern becomes unstable to the formation of an anisotropic WC at very low temperatures, breaking the particle-hole symmetry \cite{MF}. The experimental evidence for stripe phases in quantum Hall systems consists of very large anisotropies in the magnetoresistance with respect to two orthogonal directions \cite{exp}. This suggests an interpretation in terms of charge stripes, where easy electron transport along the stripe edges leads to a small resistance, but transport across the stripes involves relatively rare tunneling processes, thus explaining the large resistance in the orthogonal direction \cite{MF,OHS}. 

{\sl Quantum limit (FQHE): $d\sim 2R_C$.} 
In the regime of the phase diagram intermediate between the WC and CDW phases, the FQHE is observed in the lowest two LLs \cite{TSG}. Recent theoretical and experimental studies \cite{MS,goldman} support the description of the FQHE in terms of an incompressible liquid of CFs, each of which consists of a bound state of an electron and a vortex-like collective excitation which carries a charge of opposite sign \cite{jain}. The scaling arguments presented above show that in the quantum limit  only one length scale is present in the problem, and thus no perturbative approaches are applicable.

{\sl Effect of impurities and finite temperature.} 
The arguments presented above suggest that the observation of the FQHE
is possible in higher LLs, e.g.\ at $\bar{\nu}=1/5$ for $n=2$. However, this phase has not yet been observed, and some experiments even suggest a CDW ground state in this regime \cite{lewis}. Failure to observe the FQHE in higher LLs is likely to be due to the residual impurities in the samples, which favour crystalline structures such as the WC and CDWs \cite{jiang}. Deformation of the inhomogeneous charge structure makes these phases better adapted to follow an underlying electrostatic potential than is an incompressible, homogeneous liquid. The two transition lines therefore move towards the quantum-liquid phase, as shown by the arrows and broken lines in fig.\,\ref{fig01}. The region where the FQHE is observed can become extremely narrow in higher LLs, and may even vanish, leading to a direct crossover between the triangular CDW and WC phases with no quantum melting in the intermediate regime $d\sim 2R_C$. However, samples of higher mobility and even lower impurity concentrations are expected to reveal FQHE states in the LL $n=2$ in the future \cite{morf}.

So far we have discussed the different phases determined by the two length scales $d$ and $R_C$ only at $T=0$. Finite temperatures will introduce an additional thermal length scale defined by comparing the thermal energy $k_BT$ and the Coulomb interaction between electrons. One thus obtains $l_T=e^2/\epsilon k_BT$, which corresponds to the de\ Broglie wavelength of the free electron gas.

If $l_T\gg R_C$, {\sl i.e.} at low temperatures, thermal fluctuations may be
neglected compared to quantum fluctuations of the correlated electron liquid. 
At $l_T\sim R_C$ thermal fluctuations destroy the quantum correlations
and the FQHE disappears. In the mean-field limit, local crystalline
structure of the CDW vanishes when thermal fluctuations become important
on the length scale of the CDW periodicity $l_T\sim R_C$. This leads
to an estimate of the CDW melting temperature
$T_{CDW}(n)=Ce^2/\epsilon k_Bl_B\sqrt{2n+1}$, with $C$ a dimensionless
constant, in agreement with previous work \cite{FKS}. 
Because CDW states in high LLs are observed at relatively
low magnetic fields (decreasing LL separation), inter-LL excitations
have to be included, giving rise to a screening of the Coulomb
interaction. The dielectric constant $\epsilon$ may then be replaced
by $\epsilon(n)\sim 2n+1$ \cite{FKS,AG}. 
We stress that these scaling arguments provide an estimate
of the melting temperature for {\sl\it local} CDW order. The anisotropy observed 
at half-filling in higher LLs \cite{exp} vanishes at lower 
temperatures, indicating an isotropic distribution of local stripes, as 
proposed in a liquid-crystal picture \cite{fradkin1}.

Finite-temperature effects are most complex in the WC
phase. Minima in the longitudinal magnetoresistance, similar to the ones
arising in the FQHE regime, are experimentally observed above a
temperature $T_1$ at lowest-LL filling fractions, where the WC phase is 
expected
at $T=0$ \cite{pan}. These minima vanish above a second temperature
$T_2>T_1$. 
An estimate for $T_1$ can be obtained from the Lindemann criterion \cite{lindemann}: the WC melts when the average displacement of an electron due to thermal fluctuations is a substantial fraction of the lattice constant $\langle \Delta r^2\rangle=c_L^2 d^2$ with $c_L\sim 0.1$. Equating the potential energy for a small displacement of an electron in a WC, $e^2\langle \Delta r^2\rangle /\epsilon d^3$, to the thermal energy yields a melting temperature $T_1\approx c_L^2e^2/\epsilon k_B d$, which is independent of the magnetic field, in good agreement with recent experiments \cite{pan}. This temperature corresponds to a thermal length $l_T\sim d/c_L^2\gg l_B$. 
The liquid phase therefore exhibits quantum coherence on a length 
scale $l_T$, and may be described locally by the Laughlin wave function
for temperatures such that $l_T> l_B$, thus displaying features of the FQHE.
The scaling estimate $l_T\sim l_B$ for the
definition of a temperature $T_2$ at which this coherence is lost
turns out, however, to be rather crude because it neglects a
renormalization of the magnetic length due to CF formation in the
quantum limit. In the CF picture, the minima disappear when the
temperature reaches the activation gap \cite{MS}. This leads to the
relation $T_2\approx T_C/(2ps\pm 1)$, where $T_C$ is a constant, and
the integers $p,s$ are related to the filling factor
$\bar{\nu}=p/(2ps\pm 1)$ for the FQHE states. Comparison with experiment \cite{pan} suggests a value $T_C\sim 2$K around $\nu=1/6$. 

{\sl Conclusions.} 
We have discussed the different solid and liquid phases of quantum
Hall systems using straightforward scaling arguments. Although the
bare Coulomb potential is scale-free, the effective
potential of electrons restricted to the $n$th LL has a
range of strong interaction characterised by the cyclotron radius. The
ratio between this range and the average separation $d$ of the
electrons classifies the different phases at $T=0$. At finite
temperature, a further length scale $l_T$ enters. For $d\gg 2R_C$ a WC
is formed, which melts at a temperature $T_1$ into a quantum liquid
showing features of FQHE states. Our estimates for the
melting temperature $T_1$ show that it does not vary with $B$, in
agreement with recent experiments \cite{pan}. The quantum coherence of
the electrons is not destroyed until a higher temperature $T_2$. In
the opposite limit $d\ll 2R_C$ the mean-field approximation is
justified and predicts a CDW ground state. We present a criterion for
the validity of this approximation, which excludes CDW formation in
the lowest LLs, where the required condition cannot be satisfied
because $R_C$ coincides with the smallest length $l_B$ in the
system. In an intermediate regime $d\sim 2R_C$ quantum melting of the
crystalline structures leads to a liquid phase, which is
incompressible at certain filling factors where the FQHE is
observed. The presence of impurities reduces the FQHE regime in the
phase diagram because solid phases are better adapted to follow an
underlying impurity potential.

\vspace{-0.3cm}
\acknowledgments
\vspace{-0.3cm}
We acknowledge fruitful discussions with D.\ Baeriswyl,
L.\ Benfatto, K.\ Borejsza, P.\ Lederer, R.\ Morf, 
B.\ Normand, V.\ Pasquier, and J.\ Wakeling. This work was supported by the 
Swiss National Foundation for Scientific Research under grant No.~620-62868.00.


\begin{thebibliography}{99}

\bibitem{KvK} 
  	\Name{v. Klitzing K. {\sl et al.}}
	\REVIEW{Phys. Rev. Lett.}{45}{1980}{494}. 

\bibitem{note1}If one takes into account the Zeeman splitting of each LL, the partial filling factors of the lower and upper spin branches of the $n$th LL are, respectively,  $\bar{\nu}=\nu-2n$ and $\bar{\nu}=\nu-(2n+1)$. 

\bibitem{jiang}
	\Name{Jiang H. {\sl et al.}}
	\REVIEW{Phys. Rev. Lett.}{65}{1990}{633};
	\REVIEW{Phys. Rev. B}{44}{1991}{8107};
	\Name{Engel L. W. {\sl et al.} }
	\REVIEW{Solid State Commun.}{104}{1997}{167}.

\bibitem{TSG}
	\Name{Tsui D. C. {\sl et al.}}
	\REVIEW{Phys. Rev. Lett.}{48}{1982}{1559};
	\Name{Laughlin R. B.}
	\REVIEW{Phys. Rev. Lett.}{50}{1983}{1395}.

\bibitem{jain}
	\Name{Jain J. K.}
	\REVIEW{Phys. Rev. Lett.}{63}{1989}{199};
	\REVIEW{Phys. Rev. B}{41}{1990}{7653};
	{\sl Composite Fermions}, edited by HEINONEN O. 
	(World Scientific) 1998.

\bibitem{exp} 
	\Name{Lilly M. P. {\sl et al.}}
	\REVIEW{Phys. Rev. Lett.}{82}{1999}{394};
	\Name{Du R. R. {\sl et al.}}
	\REVIEW{Solid State Commun.}{109}{1999}{389}.

\bibitem{FKS}
	\Name{Koulakov A. A. {\sl et al.}}
	\REVIEW{Phys. Rev. Lett.}{76}{1996}{499};
	\Name{Fogler M. M. {\sl et al.}}
	\REVIEW{Phys. Rev. B}{54}{1996}{1853}.

\bibitem{MC}
	\Name{Moessner R. and Chalker J. T.}
	\REVIEW{Phys. Rev. B}{54}{1996}{5006}.

\bibitem{pan}
	\Name{Pan W. {\sl et al.}}
	\REVIEW{Phys. Rev. Lett.}{88}{2002}{176802}.

\bibitem{AG}Inter-LL excitations may be included in the random-phase
  approximation. This leads to a screened Coulomb interaction in a
  certain wave vector range but has only a minor influence on the
  physical properties of the system; see: 
	\Name{Aleiner I. L. and Glazman L. I.}
	\REVIEW{Phys. Rev. B}{52}{1995}{11296}.

\bibitem{GMP}
	\Name{Girvin S. M. {\sl et al.}}
	\REVIEW{Phys. Rev. B}{33}{1986}{2481}.

\bibitem{MS}
	\Name{Murthy G. and Shankar R.}
	cond-mat/0205326 Preprint, 2002;
	\Name{Shankar R.}
	\REVIEW{Phys. Rev. B}{63}{2001}{085322}.

\bibitem{lam}
	\Name{Lam P. K. and Girvin S. M.}
	\REVIEW{Phys. Rev. B}{30}{1984}{473}.

\bibitem{AndersonLoc}
	\Name{Abrahams E. {\sl et al.}}
	\REVIEW{Phys. Rev. Lett.}{42}{1979}{673}.

\bibitem{fogler}
	\Name{Fogler M. M. and Koulakov A. A.}
	\REVIEW{Phys. Rev. B}{55}{1997}{9326}.

\bibitem{scheidel}
	\Name{Scheidl S. and Rosenow B.}
	\REVIEW{Int. J. Mod. Phys. B}{15}{2001}{1905}.

\bibitem{MF}
	\Name{MacDonald A. H. and Fisher M. P. A.}
	\REVIEW{Phys. Rev. B}{61}{2000}{5724}.

\bibitem{OHS}
	\Name{v. Oppen F. {\sl et al.}}
	\REVIEW{Phys. Rev. Lett.}{84}{2000}{2937}.

\bibitem{goldman}
	\Name{Goldman V. J. {\sl et al.}}
	\REVIEW{Phys. Rev. Lett.}{72}{1994}{2065}.

\bibitem{lewis}
	\Name{Lewis R. M. {\sl et al.}}
	\REVIEW{Phys. Rev. Lett.}{89}{2002}{136804}.

\bibitem{morf}
	\Name{Morf R. and D'Ambrumenil N.}
	\REVIEW{Phys. Rev. Lett.}{74}{1995}{5116}.

\bibitem{fradkin1}
	\Name{Fradkin E. and Kivelson S. A.}
	\REVIEW{Phys. Rev. B}{59}{1999}{8065};
	\Name{Wexler C. and Dorsey A. T.}
	\REVIEW{Phys. Rev. B}{64}{2001}{115312}.

\bibitem{lindemann}
	\Name{Lindemann F.}
	\REVIEW{Phys. Z. (Leipzig)}{11}{1910}{69}.

\end{thebibliography}
\end{document}